\documentclass[12pt]{article}

\usepackage{amsmath,bbm,latexsym}
\usepackage{epsfig}
\usepackage{epstopdf}

\newcommand{\be}{\begin{equation}}
\newcommand{\ee}{\end{equation}}
\newcommand{\ba}{\begin{eqnarray}}
\newcommand{\ea}{\end{eqnarray}}

\begin{document}


\begin{titlepage}
\vspace*{-2cm}

\begin{center}
{\Large\bf Astrophysical Consequences of a Neutrinophilic Two-Higgs-Doublet Model
\vspace{3mm}}
\end{center}
\vskip 0.5cm
\begin{center}
{\large Marc Sher~\footnote{mtsher@wm.edu} and Christopher Triola~\footnote{ctriola@email.wm.edu}},
\vskip .3cm
\emph{High Energy Theory Group, Department of Physics,}\\
\vskip .1cm
\emph{College of William and Mary, Williamsburg, VA 23187, USA}\\
\end{center}
\vskip 0.5cm

\begin{abstract}
In a recently proposed neutrinophilic two-Higgs doublet model, the low-energy (sub-MeV) effective theory consists of a real scalar with a vev of O(0.1) eV and three Dirac neutrinos.  Other models could lead to the same low energy theory.  In this Brief Report, we study constraints on the parameter space of the model, including vacuum stability, unitarity, perturbativity and the effects on the invisible Z width.   Interestingly, we find that all neutrinos become massless at temperatures above approximately 1000 K, but can find no phenomenological effects of this finding.  The most direct test of the model is that it predicts that in a galactic supernova, the energy distributions of the electron, muon and tau neutrinos will be Fermi-Dirac with identical temperatures, unlike the conventional distributions.
 \end{abstract}
\end{titlepage}
\setcounter{footnote}{0}
\vskip2truecm

\newpage


\section{Introduction}

One of the deepest mysteries of the Standard Model concerns the nature and origin of neutrino masses.    One can easily obtain neutrino masses in the Standard Model by simply adding right-handed neutrinos and generate their masses through the conventional Higgs mechanism.   However, this requires extraordinarily small Yukawa couplings.    One can generate small masses through the seesaw mechanism, but this requires a very large mass scale for the right-handed neutrino.

An alternative model for small Dirac neutrino masses was proposed originally by Wang et al.\cite{Wang:2006jy}
and further developed by Gabriel and Nandi\cite{Gabriel:2006ns}.    In this model, a second Higgs doublet is added to the Standard Model, and a discrete symmetry couples the first doublet to the Standard Model fermions and the second to the right handed neutrinos.   By choosing a vev for the second doublet to be extremely small, of O(0.1) eV, one obtains small neutrino masses instead of extremely small Yukawa couplings.   The model contains a very light real scalar corresponding to the small vev.

Of course, the model must be very fine-tuned in order to get such a small vev.   This makes the theory quite unnatural, and certainly not more natural than simply choosing extremely small Yukawa couplings.   However, it is a logical alternative, and is no more fine-tuned than non-supersymmetric grand unified theories.     A possible mechanism for generating a small vev through higher dimensional operators and neutrino condensation is discussed in Ref. \cite{Wang:2006jy}.   We will not be concerned with the precise mechanism of generation of the small vev in this work.

In this Brief Report, we study the phenomenology of the low energy effective theory of the model, concentrating on energies well below the electron mass.   In this effective theory, the weak interactions, all charged fermions and the heavier scalar fields are all integrated out.  The only remaining fields are the three Dirac neutrinos and a single real scalar, as well as massless gauge bosons.

The Lagrangian for the model is simply
\begin{equation}
{\cal L} = \frac{1}{2}\partial^\mu\phi\partial_\mu\phi + \frac{1}{2}\mu^2\phi^2 -\frac{1}{4}\lambda\phi^4 + h_{ij}\bar{\nu}_i\nu_j\phi + {\rm H.c.}
\end{equation}

The vacuum expectation of $\phi$, $\sigma$ is given by $\sigma^2= \mu^2/\lambda$, and the mass matrix of the neutrinos is then $h_{ij}\sigma$.    Diagonalizing the neutrino mass matrix  automatically diagonalizes the Yukawa interactions.    Since the largest $\Delta m^2$ in the neutrino sector \cite{Amsler:2008zzb} is approximately $(0.05\  {\rm eV})^2$, and since the Yukawa coupling cannot be too much larger than $1$, one expects $\sigma$ to be $O(0.1)$ eV.     Note that this Lagrangian is obtained by integrating out the heavy fields from the original Lagrangian of the Two-Higgs-Doublet model discussed above, so the couplings to $W$ bosons, electrons and other standard model particles are suppressed by factors of $\sigma^2/M^2_W$.   The $\phi$ and $\nu$'s are thus effectively gauge singlets in the effective theory.

\section{Constraints on the Parameters and Phenomenology}

Some bounds on these parameters can be obtained by considering unitarity, by requiring that the theory be stable and perturbative from the eV scale to the scale at which the effective theory is no longer valid and that the contribution to the Z width be phenomenologically acceptable.

The unitarity bound is obtained by expanding the matrix element in $\phi\phi \rightarrow \phi\phi$ in partial waves, and using the optical theorem to show that the coefficients in the partial wave expansion are bounded by $|\Re(a_\ell)| < \frac{1}{2}$.    The strongest bound comes from the $\ell=0$ partial wave, and gives $\lambda < 4\pi/3$.    

\begin{figure}[ht]
\begin{center}
\includegraphics[width=12cm, height=15cm]{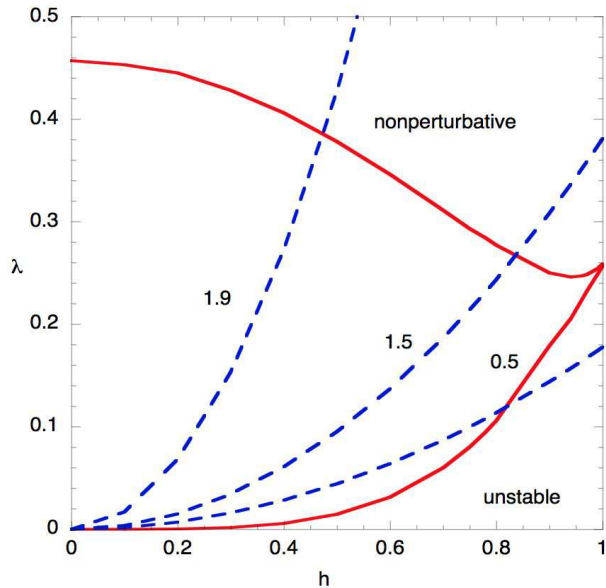}
\vskip -3cm
\caption{Allowed region of parameter space.   The dotted lines give the critical temperature divided by $\sigma$.    As discussed in Section 3, above the critical temperature, the vev of $\phi$ vanishes and neutrinos become massless.}
\end{center}
\end{figure}

Since this is an effective theory which must be valid from the sub-eV scale to the electron mass  scale, the potential must be bounded and perturbative throughout that range.    The RGEs for the above potential are given by \cite{Sher:1988mj}
\begin{eqnarray}
\frac{d\lambda}{dt} & = & \frac{1}{8\pi^2}(9\lambda^2+6\lambda h^2 - h^4) \\
\frac{dh}{dt} & = & \frac{9}{32\pi^2}h^3 
\end{eqnarray}
where $t=\ln{M/\sigma}$, where $M$ is the renormalization scale and one integrates from $t=0$ to $t= \ln{m_e/\sigma} \sim 17$.   The choice of $m_e$ to cut off the integration is, of course, somewhat arbitrary; the exact cutoff depends on the ultraviolet completion of the theory.         Integrating these equations leads to the bounds in Figure 1.       A given point in the allowed region corresponds to allowed values of the neutrino mass and the scalar mass.

One can consider the effects of the additional states on the invisible Z-width.   The processes $Z\rightarrow \bar{\nu}\nu \phi\phi$ and $Z \rightarrow \bar{\nu}\nu\phi$ were discussed in Ref. \cite{Gabriel:2006ns} and shown to be negligible.    One can also consider $Z\rightarrow \bar{\nu}_R\nu_R$, which can occur at one-loop through $\phi$ exchange.   The effective number of neutrinos depends on the details of the theory in the ultraviolet; using a cutoff of $400$ ($1000$) GeV gives $0.0004 h^4$ ($0.0026 h^4$).    Given that the experimental error in the effective number of neutrinos\cite{Amsler:2008zzb}  is $0.008$, this gives no bound in the allowed region of Figure 1.   In addition, the effects of the new light states on nucleosynthesis are discussed in Ref. \cite{Wang:2006jy} and shown to be acceptable.

A clear prediction of this model is that there will be a contribution to the invisible decay of the Standard Model Higgs boson ($h \rightarrow \phi\phi$), whose size depends sensitively on the value of the scalar self-couplings.    In addition, effects of the charged Higgs and pseudoscalar (which depend, of course, on the ultraviolet completion of the model) are discussed in Refs. \cite{Wang:2006jy,Gabriel:2006ns}.

Other than the result in Figure 1, most of the above is discussed in Refs. \cite{Wang:2006jy,Gabriel:2006ns}.   They also point out that the mixing between the light $\phi$ field and all heavy fields is $O(\sigma^2/M^2_W)$, and thus contributions to electroweak precision measurements, such as the $\rho$ parameter, are negligible.  In this Report, we wish to point out some additional effects that have not yet been discussed, and which can have dramatic implications.

\section{Astrophysical Effects}

One of surprising consequences of the potential in Equation 1 are the effects of finite temperature.   As in the Standard Model, at high temperature a quadratic term $T^2\phi^2$ is added, resulting in $\langle\phi\rangle\rightarrow 0$ at sufficiently high temperature.    The inclusion of high temperature terms is straightforward \cite{Sher:1988mj}, and the critical temperature at which the vev of $\phi$ vanishes is 
\begin{equation}
\frac{T_c}{\sigma} = \sqrt{\frac{24\lambda}{6\lambda + 4h^2}}
\end{equation}
In Figure 1, we have plotted the value of $\frac{T_c}{\sigma}$ for various points in the $(\lambda,h)$ plane.  One can see that $0.5 < T_c/\sigma < 2.0$ for all of the allowed parameter space.  

For the normal hierarchy, the mass of the heaviest neutrino is $0.05$ eV , and thus $\sigma$ is between $0.1$ and $0.5$ eV for most of the allowed parameter space.    This means that the critical temperature will be between $0.05$ eV and $1.0$ eV.    The rather surprising result is that the vev of $\phi$, and thus all neutrino masses, will vanish at temperatures between $550$ and $11000$ Kelvin!     This would mean that neutrino oscillations vanish at relatively low temperatures.

It is important to remember, however, that ``temperature"   refers to the {\it neutrino} temperature (or, equivalently, the $\phi$ temperature), not the photon temperature.    In the center of the Sun, for example, the temperature of the photons, electrons and hadrons is typically several million Kelvin.     However, these states have extremely weak interactions with the $\nu$ and $\phi$ fields, and thus will not ``heat" up the $\phi$ field quickly.    Since the coupling is typically of the order of $10^{-12}$ \cite{Wang:2006jy,Gabriel:2006ns}, the equilibrium time will be extremely long,  of $O(10^{24})$ times longer than typical weak scale decays.     The motion of the Sun through the background $\phi$ field will be too rapid for any equilibration to take place.    Thus solar neutrinos will still oscillate, as will neutrinos generated in accelerator experiments.

There is an additional effect of this model.    Neutrinos will be able to interact with each other via exchange of a $\phi$.   This will be relatively strong, and on distance scales smaller than the Compton wavelength of the $\phi$ will be as strong as a typical electromagnetic interaction.  The Compton wavelenghth of the $\phi$ will typically be of O(10) microns.     Thus neutrinos will interact with electromagnetic strength on distance scales smaller than O(10) microns!    Is there any way to detect this?   A typical neutrino beam is much larger than this distance, so there would not be any substantial focusing from a pinch effect, and focusing on scales smaller than this does not appear to be detectable.

Where could these effects be seen?     The only times that neutrinos are in equilibrium at a high temperature are during the big bang and in supernovae.      A small neutrino mass in the big bang could have an effect on the relic neutrino density, but the relic neutrinos have not been detected.   Thus, the only hope of seeing the vanishing of neutrino masses at high temperature is in supernovae.   What about the strong interaction between neutrinos (which exists even at low temperatures)?     The $\phi$ field could be emitted copiously in stars, of course.  However,  the $\phi$ field would likely be heavier than some of the neutrinos, and would thus decay into them.    A rough estimate shows that they would not appreciably escape from stars.    Again, the only possible way to measure this effect is in supernovae.

There have been hundreds of studies of supernovae neutrinos in the past decade.    A review with a large number of references can be found in \cite{Raffelt:2007nv} and a recent PhD thesis by Esteban-Pretel with a very clear discussion of neutrino energies and spectra is in Ref. \cite{EstebanPretel:2009ca}.     An important effect that was neglected until fairly recently is the MSW-type collective effect with other neutrinos in the neutrinosphere (see Ref. \cite{Duan:2010bg} for a review).    There are three energy distributions of neutrinos in a supernova---the $\nu_e$, $\bar{\nu}_e$ and $\nu_x$ energy distributions, where $\nu_x$ refers to muon and tau neutrinos and antineutrinos (which will be the same).     One expects the average energies, $\epsilon$,  to differ somewhat, with $\epsilon_{\nu_x} > \epsilon_{\bar{\nu}_e} > \epsilon_{\nu_e}$, and the energy distributions will differ somewhat from a Fermi-Dirac distribution (see Ref. \cite{EstebanPretel:2009ca} for a discussion of the difference distributions used).     Although current detectors would only be sensitive to $\bar{\nu}_e$, future detectors of a galactic supernova may be sensitive to all of the distributions \cite{Dasgupta:2011wg}.

In this model, the biggest effect will be the strong interactions between neutrinos.   The $\phi$ will mediate $\bar{\nu}_i\nu_i\leftrightarrow \bar{\nu}_j\nu_j$ interactions with electromagnetic strength.    This will result in the complete equilibration of all of the neutrino species, resulting in pure Fermi-Dirac distributions of identical temperatures.    As a result, issues of neutrino oscillations in the neutrinosphere and collective effects will be completely washed out.    A prediction of this model is that the neutrino energies and distributions of the various neutrino and antineutrino species will be identical, in sharp contrast to the standard picture.     Note that the effect discussed earlier of the neutrinos becoming massless at high temperature can thus not be detected.

\section{Conclusion}

In a neutrinophilic two-Higgs-doublet model, the effective low-energy theory below the MeV scale consists of three Dirac neutrinos and a light scalar.    This may be the low-energy limit of other models as well.   We have studied the allowed parameter-space of the model, and have found two remarkable properties.   Neutrinos will become absolutely massless at temperatures above a few thousand Kelvin, and there will be electromagnetic-strength interactions between neutrinos at distance scales below O(10) microns.     However, the only phenomenological evidence for such properties would occur in the detection of neutrinos from a galactic supernova, in which the model predicts (in contrast to the standard picture) that all of the neutrino species will have Fermi-Dirac distributions of identical temperatures.    This would be true of any model with extremely strong neutrino-neutrino interactions.

We are very grateful to Gail McLaughlin and Jim Kneller for several discussions about the properties of supernova neutrinos, and Chris Carone for reading the manuscript.  This work was supported by the National Science Foundation  grant NSF-PHY-0757481.


\end{document}